\documentclass[twocolumn,english,aip,reprint,superscriptaddress]{revtex4}
\usepackage[T1]{fontenc}
\usepackage[latin9]{inputenc}
\usepackage{verbatim}
\usepackage{units}
\usepackage{amsmath}
\usepackage{amssymb}
\usepackage{graphicx}
\usepackage{esint}

\makeatletter
\@ifundefined{textcolor}{}
{%
 \definecolor{BLACK}{gray}{0}
 \definecolor{WHITE}{gray}{1}
 \definecolor{RED}{rgb}{1,0,0}
 \definecolor{GREEN}{rgb}{0,1,0}
 \definecolor{BLUE}{rgb}{0,0,1}
 \definecolor{CYAN}{cmyk}{1,0,0,0}
 \definecolor{MAGENTA}{cmyk}{0,1,0,0}
 \definecolor{YELLOW}{cmyk}{0,0,1,0}
}

\usepackage[english]{babel}
\addto\captionsenglish{%
}

\usepackage[none]{hyphenat}

\usepackage{babel}

\makeatother

\usepackage{babel}
\begin{document}

\title{Topological Charge Analysis of Ultrafast Single Skyrmion Creation}

\author{Gen Yin}

\affiliation{Department of Electrical and Computer Engineering, University of California, Riverside,
CA 92521-0204, USA}

\author{Yufan Li}

\affiliation{Department of Physics and Astronomy, Johns Hopkins University, Baltimore,
MD 21218, USA}

\author{Lingyao Kong}

\affiliation{State Key Laboratory of Surface Physics and Department of Physics,
Fudan University, Shanghai 200443, China}

\author{Roger K. Lake}

\thanks{Author to whom correspondence should be addressed. email: rlake@ee.ucr.edu}

\affiliation{Department of Electrical and Computer Engineering, University of California, Riverside,
CA 92521-0204, USA}

\author{Chia-Ling Chien }

\affiliation{Department of Physics and Astronomy, Johns Hopkins University, Baltimore,
MD 21218, USA}

\author{Jiadong Zang}

\thanks{Author to whom correspondence should be addressed. email: jiadong.zang@unh.edu}

\affiliation{Department of Physics, University of New Hampshire, Durham, NH 03824,
USA}
\begin{abstract}
Magnetic skyrmions are topologically non-trivial spin textures of potential interest
for future information storage applications, and for such purposes, the control and understanding of
single skyrmion creation is required. 
A scheme is analyzed to create single N\'{e}el-type and Bloch-type
skyrmions in helimagnetic
thin films utilizing the dynamical excitations induced by the Oersted
field and the spin transfer torque given by a vertically injected spin-polarized current. 
A topological charge analysis using a lattice version of the topological charge 
provides insight into
the locally triggered transition from a trivial to a non-trivial topological
spin texture of the N\'{e}el or Bloch type skyrmion.
The topological protection 
of the magnetic skyrmion 
is determined by the symmetric Heisenberg exchange energy. 
The critical switching current density is $\sim10^{7}\thinspace\textrm{A/cm}^{2}$,
which decreases with the easy-plane type uniaxial anisotropy and
thermal fluctuations. 
In-plane spin polarization of the injected current performs better than 
out-of-plane polarization, and it
provides ultrafast switching times (within 100 ps)
and reliable switching outcomes. 
\end{abstract}
\maketitle

\section{Introduction}

Magnetic skyrmions are topologically protected spin textures in which
the local moments on a two dimensional lattice point in all directions
with a topologically nontrivial mapping to the unit sphere \cite{Nagaosa_NatNano_Review,rosler_spontaneous_2006}.
Physically, each skyrmion is a circular spin texture in which the
spins on the periphery are polarized vertically, the central spin
is polarized in the opposite direction, and, in between, the spins
smoothly transition between the two opposite polarizations. 
A swirling
transition, which is effectively a circle of double Bloch-type domain
wall, gives a Bloch-type skyrmion. 
This type of skyrmion was first
discovered in the temperature-magnetic field (T-H) phase diagram of
B20 magnets \cite{muhlbauer_skyrmion_2009,yu_real-space_2010,yu_near_2011}.
In these materials, the atomic structure of the lattice breaks the
inversion symmetry, inducing an asymmetric Dzyaloshinsky-Moriya (DM)
exchange interaction \cite{dzyaloshinsky_thermodynamic_1958,moriya_anisotropic_1960}.
The competition between the DM exchange and the symmetric Heisenberg
exchange stablizes the skyrmion phase in these materials. 
A N\'{e}el-type
skyrmion, on the other hand, is a wrapped double N\'{e}el-wall. 
Such a
skyrmion is stabilized by an interfacial DM interaction, which is originated
from the broken interfacial inversion symmetry. %
This type of DM interaction
is usually observed at the interface between a magnetic thin film
and a layer of heavy metal with strong spin-orbit coupling (SOC). 
For both types of skyrmions, the radius, ranging from about 3 nm to
100 nm \cite{kanazawa_large_2011-1,muhlbauer_skyrmion_2009,yu_near_2011,yu_real-space_2010},
is determined by the ratio of the strengths of the DM interaction
and the Heisenberg interaction \cite{han_skyrmion_2010}. 
Skyrmion lattices and isolated skyrmions in both bulk and thin films
have been observed by neutron scattering \cite{muhlbauer_skyrmion_2009,munzer_skyrmion_2010},
Lorentz transmission electron microscopy \cite{yu_real-space_2010,yu_near_2011,seki_observation_2012,yu_skyrmion_2012,li_robust_2013},
and spin-resolved scanning tunneling microscopy (STM) \cite{heinze_spontaneous_2011}. 
Current can drive skyrmion spin textures with a current density 4-5
orders of magnitude lower than that required to move conventional
magnetic domain walls \cite{jonietz_spin_2010,schulz_emergent_2012,yu_skyrmion_2012,zang_dynamics_2011}. 
This suggests promising spintronic applications exploiting the topological
spin texture as the state variable
\cite{fert_skyrmions_2013,sampaio_nucleation_2013,iwasaki_current-induced_2013}. 
A two-dimensional skyrmion lattice may be formed under a uniform magnetic
field \cite{muhlbauer_skyrmion_2009,yu_real-space_2010}, however,
the switching of isolated, individual skyrmions is far more challenging. 
The single skyrmion switching was first experimentally demonstrated
by injecting spin-polarized current from an STM tip into ultra-thin
Pd/Fe/Ir(111) films at 4.2~K \cite{romming_writing_2013}. 
Other
schemes of single skyrmion switchings, such as using a sharp notch \cite{iwasaki_current-induced_2013},
a circulating current \cite{tchoe_skyrmion_2012}, thermal excitations \cite{finazzi_laser-induced_2013}
and spin-orbit torques (SOTs) \cite{sampaio_nucleation_2013} have
been proposed. 
Spintronic applications call for on-wafer solutions
to precisely control the position and the time of skyrmion switchings
with good reliability. 
 This is rather difficult because each switching
event corresponds to a topological transition, which has to break
the protection given by the topological order. 
This process has to
overcome the topological protection barrier, which is both energetically
unfavorable and difficult to manipulate. 

In this paper, we theoretically investigate the topological transition
of the microscopic spin texture during a dynamical skyrmion creation
process. 
 This picture provides insight into the critical condition
to create isolated skyrmions and a quantitative understanding in the
topological barrier. 
Based on this picture, we propose that controlled
skyrmion creation can be realized by the spin transfer torques (STTs)
induced from a magnetic electrode. 
Such a geometry and creation mechanism
is applicable to both the Bloch-type and the N\'{e}el-type skyrmions,
and is potentially compatible with the standard metal process used
in silicon integrated circuits.

\section{Topological transition analysis}

The critical condition of the topological transition is determined
by the topological charge evolution during a spin dynamical process. 
A skyrmion is distinguished from a ferromagnet or other trivial state
by the topological charge $Q$, which is a nonvanishing integer \cite{skyrme_non-linear_1961,rajaraman_solitons_1987}. 
Each skyrmion contributes $\pm1$ to the total topological charge. 
Usually $Q=\frac{1}{4\pi}\int d^{2}r\mathbf{S}\cdot(\partial_{x}\mathbf{S}\times\partial_{y}\mathbf{S})$
is employed, but it is well defined only in the continuum limit where
all the spins are almost parallel to their neighbors \cite{rajaraman_solitons_1987}. 
In this limit, magnetic dynamical processes can only distort the geometry
of the spin texture, but cannot change the wrapping number in the
spin space.  
Thus, the topological charge above is conserved, and can
not capture the precise time evolution of the topological transition.
\begin{figure}
\centering{}\includegraphics[width=1\columnwidth]{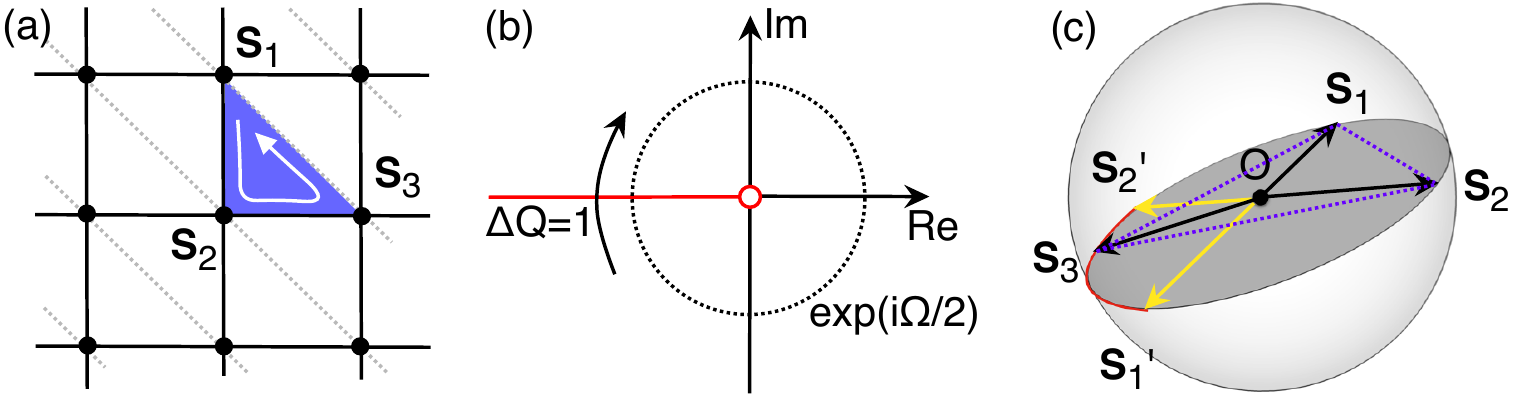}
\protect\protect\caption{(color online) The critical conditioon of a topological transition.
(a), Triangulated square lattice. $\mathbf{S}_{1}$, $\mathbf{S}_{2}$
and $\mathbf{S}_{3}$ follow a counter-clockwise order on each triangle
grid. (b), $\exp\left(\frac{i\Omega}{2}\right)$ on the complex plane.
The branch cut is denoted by the red line on the negative real axis.
(c), A typical spin configuration at the moment of a topological transition.
It only occurs when $\mathbf{S}_{3}$ crosses the geodesic $\mathbf{S}_{1}'\mathbf{S}_{2}'$
(red arc). $\mathbf{S}_{1}'$ and $\mathbf{S}_{2}'$ are the point
reflection images of $\mathbf{S}_{1}$ and $\mathbf{S}_{2}$ about
the sphere center. \label{fig:BrachCut_CriticalCondition}}
\end{figure}

Here we employ the lattice version of the topological charge that
provides a microscopic picture of the spin evolution and reveals the
microscopic criteria for a topological transition to occur during
any dynamical process. 
 This version of $Q$ was first introduced by
Berg \emph{et al.} \cite{berg_definition_1981}, which is defined on
a square lattice mesh illustrated in Fig. \ref{fig:BrachCut_CriticalCondition}
(a). 
 The calculation of $Q$ starts by triangulating the entire lattice
and then counting the solid angles $\Omega_{\bigtriangleup}$ for
each triangle $\bigtriangleup(\mathbf{S}_{1},\mathbf{S}_{2},\mathbf{S}_{3})$
determined by 
\begin{equation}
\exp(i\frac{\Omega_{\bigtriangleup}}{2})=\rho^{-1}[1+\mathbf{S}_{1}\cdot\mathbf{S}_{2}+\mathbf{S}_{2}\cdot\mathbf{S}_{3}+\mathbf{S}_{3}\cdot\mathbf{S}_{1}+i\mathbf{S}_{1}\cdot(\mathbf{S}_{2}\times\mathbf{S}_{3})],\label{eq:triangleCharge}
\end{equation}
where $-2\pi<\Omega<2\pi$ and $\rho=[2(1+\mathbf{S}_{1}\cdot\mathbf{S}_{2})(1+\mathbf{S}_{2}\cdot\mathbf{S}_{3})(1+\mathbf{S}_{3}\cdot\mathbf{S}_{1})]^{1/2}$
is the normalization factor \cite{berg_definition_1981}. The lattice
version of the topological charge $Q$ is then given by summing over
all of the triangles. 
\begin{equation}
Q=\frac{1}{4\pi}\sum_{\triangle}\Omega_{\triangle}\label{eq:topologicalCharge}
\end{equation} 
From this definition, the directional solid angle $\Omega_{\bigtriangleup}$
ranges from $-2\pi$ to $2\pi$ so that the negative real axis of
the complex plane in Eq. (\ref{eq:triangleCharge}) is a branch cut. 
The exponential $e^{i\Omega_{\bigtriangleup}/2}$ lies on the branch
cut in the complex plane when $\mathbf{S}_{1}\cdot(\mathbf{S}_{2}\times\mathbf{S}_{3})=0$,
and $1+\mathbf{S}_{1}\cdot\mathbf{S}_{2}+\mathbf{S}_{2}\cdot\mathbf{S}_{3}+\mathbf{S}_{3}\cdot\mathbf{S}_{1}<0$.
$\Omega_{\bigtriangleup}$ is $2\pi$ immediately above, and $-2\pi$
immediately below, the branch cut. 
 Any dynamical process causing $e^{i\Omega_{\bigtriangleup}/2}$
to cross the branch cut is accompanied by a change in the topological
charge of $\pm1$ as shown in Fig. \ref{fig:BrachCut_CriticalCondition}
(b). 
 To trigger an event crossing the branch cut, the dynamical process
must drive three spins $\mathbf{S}_{1}$, $\mathbf{S}_{2}$, $\mathbf{S}_{3}$
in one particular triangle coplanar from the condition $\mathbf{S}_{1}\cdot(\mathbf{S}_{2}\times\mathbf{S}_{3})=0$. 
The other condition $1+\mathbf{S}_{1}\cdot\mathbf{S}_{2}+\mathbf{S}_{2}\cdot\mathbf{S}_{3}+\mathbf{S}_{3}\cdot\mathbf{S}_{1}<0$
leads to the inequality $(\mathbf{S}_{1}-\mathbf{S}_{2})\cdot(\mathbf{S}_{3}-\mathbf{S}_{2})>0$,
so that $\angle\mathbf{S}_{1}\mathbf{S}_{2}\mathbf{S}_{3}$ is an
acute angle, and the same holds true for permutations of the three
indices $1$, $2$, and $3$. 
 Consequently, three spins must point
`away' from each other at the branch cut. For fixed $\mathbf{S}_{1}$
and $\mathbf{S}_{2}$, $\mathbf{S}_{3}$ must lie on the arc ${\bf S}_{1^{\prime}}{\bf S}_{2^{\prime}}$
as shown in Fig. \ref{fig:BrachCut_CriticalCondition} (c). 
 This coplanar
but highly non-collinear critical state must be achieved during skyrmion
switching events. 
Based on this switching criteria, the energy barrier protecting the
topological charge is identified, and can thus be quantified. 
The full spin Hamiltonian of a magnetic helix is given by 
\begin{equation}
H=\sum_{\left\langle i,j\right\rangle }[-J\mathbf{S}_{i}\cdot\mathbf{S}{}_{j}+H_{i,j}^{\textrm{DM}}]-\mu_{0}\sum\mathbf{S}_{i}\cdot\mathbf{H}_{\textrm{Ost}},
\end{equation}
where 
\[
\begin{cases}
H_{i,j}^{\textrm{DM}}=D\mathbf{\hat{r}}_{ij}\cdot\left(\mathbf{S}\times\mathbf{S}_{j}\right) & \left(\textrm{Bloch type}\right)\\
H_{i,j}^{DM}=D\left(\hat{\mathbf{z}}\times\hat{\mathbf{r}}_{ij}\right)\cdot\left(\mathbf{S}_{i}\times\mathbf{S}_{j}\right) & \left(\textrm{N\'{e}el type}\right)
\end{cases}.
\]
The two terms in the square bracket are the Heisenberg and DM interactions,
respectively, and the last term is the Zeeman coupling. 
 $\hat{r}_{ij}$
denotes the unit vector pointing from $\mathbf{S}_{i}$ to $\mathbf{S}_{j}$. 
At the moment of switching,
three spins on one particular triangle are coplanar, and the DM interaction
does not contribute to the total energy. 
 The energy of this particular
triangle, measured from the ferromagnetic state, is thus given by
\begin{align}
\Delta\epsilon  = &
J\left(1-\frac{\mathbf{S}_{1}\cdot\mathbf{S}_{2}}{2} - 
\frac{\mathbf{S}_{2}\cdot\mathbf{S}_{3}}{2}\right)
\nonumber
\\
& + \mathbf{B}_{\textrm{Ost}} \cdot \left(\frac{1}{2}-\frac{\mathbf{S}_{1}+\mathbf{S}_{2}+\mathbf{S}_{3}}{6}\right).
\label{eq:Barrier}
\end{align}
Since the spins at the transition are highly non-collinear, the exchange
becomes very large, and the Zeeman coupling in the second term in
Eq. (\ref{eq:Barrier}) is thus negligible. 
From the topological transition
requirement, 
$1+\mathbf{S}_{1}\cdot\mathbf{S}_{2}+\mathbf{S}_{2}\cdot\mathbf{S}_{3}+\mathbf{S}_{3}\cdot\mathbf{S}_{1}<0$,
it can be obtained that 
$-J(\mathbf{S}_{1}\cdot\mathbf{S}_{2}+\mathbf{S}_{2}\cdot\mathbf{S}_{3})>J(1+\mathbf{S}_{3}\cdot\mathbf{S}_{1})\geq0$.
Thus, $\Delta\epsilon>J$ has to be satisfied. 
The maximum value of
$\Delta\epsilon=2J$ occurs when $\mathbf{S}_{2}$ is anti-aligned
with both $\mathbf{S}_{1}$ and $\mathbf{S}_{3}$, such that $J<\Delta\epsilon<2J$.
In different switching processes, the actual value of this barrier
varies within this range, determined by the exact spin configurations
at the moment of the transition. 
Since this criteria comes from the
generic topological charge analysis, it applies for both the Bloch-type
and the N\'{e}el-type skyrmions. 
\section{Single skyrmion creation due to a vertical current}

\begin{figure}
\begin{centering}
\includegraphics[width=1\columnwidth]{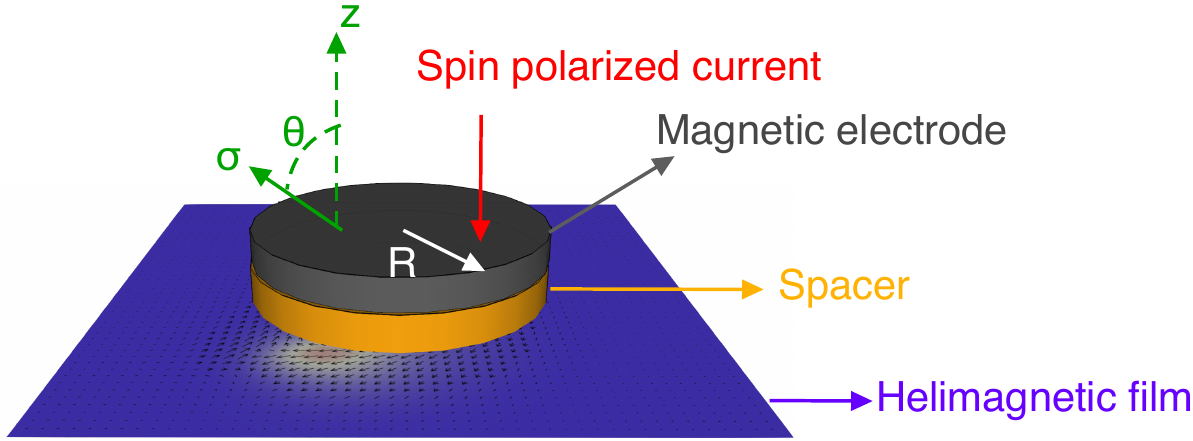} 
\par\end{centering}

\protect\protect\caption{The sandwich structure of the proposed skyrmion creation scheme.
$\theta$ is the angle between the injected spin, $\boldsymbol{\sigma}$,
and the $z$ axis.}
\end{figure}

Our proposed scheme to control the precise location and the moment
of the topological transition is illustrated in Fig. \ref{fig:DeviceSetup_SnapShots}
(a). 
A spin-less metallic (usually copper) nanopillar electrode of
radius $R$ is deposited on top of a helimagnetic thin film, with
a back contact on the bottom of the film which serves as the drain
of the electron current. 
In order to polarize the injected current,
a magnetic layer is deposited on top of the copper spacer. 
The angle
between the polarization and the x-y plane is $\theta$. 
A uniform
external magnetic field ${\bf H}_{0}$ is always applied vertically
to ensure a ferromagnetic ground state in which all spins are perpendicularly
polarized. 
In order to quantitatively evaluate the required condition
and the feasibility, dynamical simulations of the spin system based
on the Landau-Lifshitz-Gilbert (LLG) equation are performed. 
The equiation
of motion is given by 
\begin{equation}
\mathbf{\dot{S}}=-\gamma\mathbf{S}\times\mathbf{H}_{{\rm eff}}+\alpha\mathbf{S}\times\mathbf{\dot{S}}+\boldsymbol{\tau}_{\textrm{STT}}\label{eq:LLG}
\end{equation}
where $\gamma=g/\hbar$ is the gyromagnetic ratio and $\alpha$ is
the Gilbert damping coefficient. $\mathbf{H}_{{\rm eff}}$ is the
effective field given by $\mathbf{H}_{{\rm eff}}=-\partial H/\partial\mathbf{S}$. 
A fourth order Runge-Kutta algorithm is employed to integrate this
first order differential equation. 
In our simulation, material parameters
of FeGe are applied, such that $J=aA_{0}$ and $D=a^{2}D_{0}$, where
$a=2.3\thinspace\textrm{nm}$ is the choice of the mesh grid size,
$A_{0}=5.33\thinspace\textrm{meV\AA}^{-1}$ is the exchange stiffness
and $D_{0}=0.305\thinspace\textrm{meV\AA}^{-2}$ is the DM interaction
density. 
 These parameters are chosen such that the simulated helical
state period matches with the experimental observation $\lambda=2\pi a/\arctan\left(D/\sqrt{2}J\right)=70\thinspace\textrm{nm}$. 
The STT term is written as $\tau_{\textrm{STT}}=-j\frac{\gamma\hbar p}{2e\mu_{0}M_{s}t}\left[\mathbf{S}\times\left(\mathbf{S}\times\boldsymbol{\sigma}\right)\right]$
\cite{slonczewski_current-driven_1996}, where $p$ is the polarization,
$j$ is the current density, $\boldsymbol{\sigma}$ is the injected
spin orientation, $M_{s}=10^{5}\thinspace\textrm{Am}^{-1}$ is the
saturation magnetization and $t$ is the film thickness. 
 A background
field, $\mathbf{H}_{0}$, is applied along the $\hat{z}$ direction,
perpendicular to the thin film, such that the energy of a FM state
matches the energy of a single skyrmion.

\subsection{Oersted field induced creation}

\begin{figure}
\begin{centering}
\includegraphics[width=1\columnwidth]{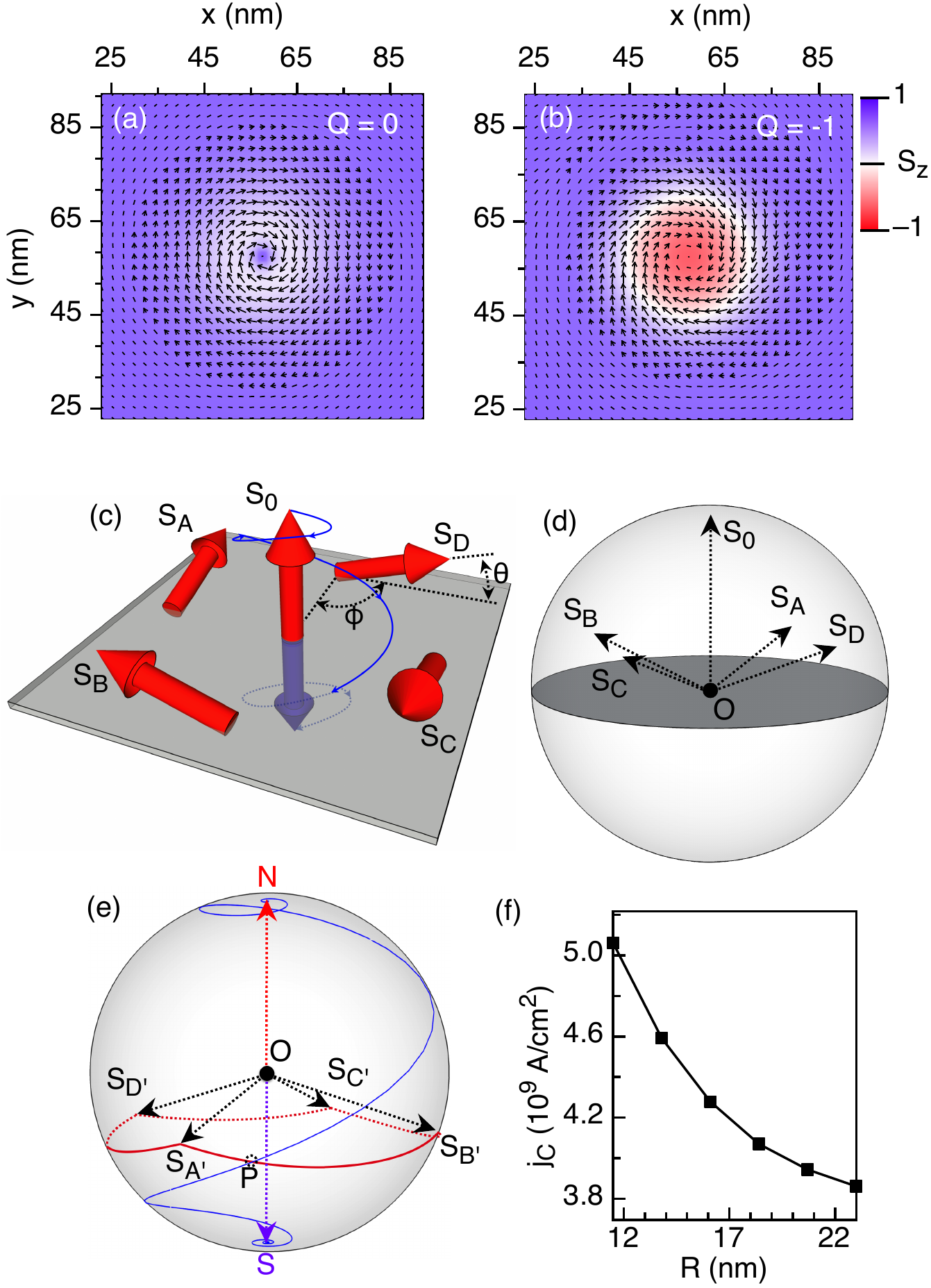} 
\par\end{centering}

\protect\protect\caption{(color online) Single Bloch-type skyrmion creation due to an Oersted
field induced by spin unpolarized current. (a) The swirling texture
before the creation of a skyrmion. (b) The spin texture after a topological
charge of $-1$ is created. (c) The real-space spin configuration
at the center of the swirling texture, immediately before the creation
moment. (d) The spin configuration in spin space. (e) The trajectory
of $\mathbf{S}_{0}$ during the creation process. The closed loop
$\mathbf{S}_{A'}\mathbf{S}_{B'}\mathbf{S}_{C'}\mathbf{S}_{D'}$ illustrates
the boundary of the topological transition. (f) The critical current
density, $j_{C}$, at different values of the electrode radius. \label{fig:Oersted_creation}}
\end{figure}

First, we consider the creation of a Bloch-type skyrmion
by the injection of spin {\em unpolarized} current,
where all the excitations in the spin texture are induced by the Oersted
field associated with the vertical current. 
Starting from a ferromagnetic (FM) initial state, an unpolarized
DC current is injected at $t=0$. 
This generates a swirling Oersted
field in the plane of the helimagnetic thin film, dragging the spins
into a swirling spin texture, which eventually evolve to a single
skyrmion at the center. 
The spin textures before and after this topological
transition are shown in Fig. \ref{fig:Oersted_creation} (a) and (b). 
At the center of the swirling texture, the central spin, $\mathbf{S}_{0}$,
and its four nearest neighbors $\mathbf{S}_{A}$, $\mathbf{S}_{B}$,
$\mathbf{S}_{C}$, and $\mathbf{S}_{D}$ form a configuration illustrated
in Fig. \ref{fig:Oersted_creation} (c). 
Due to the rotational symmetry
of the applied field, these four spins relate to each other by successive
rotations of $\pi/2$ about the $\hat{z}$ axis. 
They thus share the
same angle $\theta$ to the plane of the film, and the same azimuthal
angle $\varphi$ measured from the $x$ or $y$ axis, respectively. 
In the case of a Bloch-type skyrmion, the effective field experienced
by the central spin is along the $z$ direction with an amplitude
of $H_{{\rm eff}}^{0}=4J\sin\theta-4D\cos\theta\sin\varphi+H_{0}$
where $J$ and $D$ are the strength of the Heisenberg and the DM
interaction respectively. 
The direction of the electrical current
is chosen so that the the swirling direction of the circulating field
is the same as that of the in-plane spin component of a skyrmion;
therefore $\varphi$ is about $\pi/2$, and $\sin\varphi$ is positive. 
Before a skyrmion can be created, the circulating field pulls spins
$\mathbf{S}_{A}$, $\mathbf{S}_{B}$, $\mathbf{S}_{C}$, and $\mathbf{S}_{D}$
downward towards the plane reducing the angle $\theta$. 
$H_{{\rm eff}}^{0}$
therefore decreases accordingly, but still remains positive. 
When $\theta$ reaches a critical threshold as the four spins rotate towards
the plane, $H_{{\rm eff}}^{0}$ reverses its sign, and as a result,
spin $\mathbf{S}_{0}$ quickly flips down into the $-z$ direction. 
This process changes the topological charge by an integer and thus
creates a skyrmion. 
Note that the contribution from the DM interaction
in N\'{e}el-type skyrmions cannot generate a negative term in $H_{\textrm{eff}}^{0}$,
therefore the swirling Oersted field can only create a Bloch-type
skyrmion. 

To demonstrate this process, we draw $\mathbf{S}_{0}$ and its nearest
neighbors, $\mathbf{S}_{A}$, $\mathbf{S}_{B}$, $\mathbf{S}_{C}$,
and $\mathbf{S}_{D}$, in a unit sphere at the critical state immediately
before the reversal of the central spin {[}Fig. \ref{fig:Oersted_creation}
(d){]}. 
$\mathbf{S}_{A^{\prime}}$, $\mathbf{S}_{B^{\prime}}$, $\mathbf{S}_{C^{\prime}}$
and $\mathbf{S}_{D^{\prime}}$ are the mirror points of $\mathbf{S}_{A}$,
$\mathbf{S}_{B}$, $\mathbf{S}_{C}$, and $\mathbf{S}_{D}$ with respect
to the sphere center {[}Fig. \ref{fig:Oersted_creation} (e){]}. 
Both planes $\mathbf{S}_{A}\mathbf{S}_{B}\mathbf{S}_{C}\mathbf{S}_{D}$
and $\mathbf{S}_{A^{\prime}}\mathbf{S}_{B^{\prime}}\mathbf{S}_{C^{\prime}}\mathbf{S}_{D^{\prime}}$
are parallel with the equatorial plane, and the four points in each
plane are equidistant. 
Through a fast process, $\mathbf{S}_{0}$ rapidly
switches from the north pole ($N$) to the south pole ($S$). 
When $\mathbf{S}_{0}$ is located on the geodesic arc 
$\mathbf{S}_{A^{\prime}}\mathbf{S}_{B^{\prime}}$
shown as point $P$ in Fig. \ref{fig:Oersted_creation} (e), the three
spins $\mathbf{S}_{A}$, $\mathbf{S}_{B}$, and $\mathbf{S}_{0}$
are coplanar. 
As $\mathbf{S}_{0}$ crosses arc $\mathbf{S}_{A^{\prime}}\mathbf{S}_{B^{\prime}}$,
the solid angle formed by the three spins changes sign resulting in
a change in $\Omega_{\bigtriangleup}$ of $4\pi$ and a change in
the topological charge in Eq. (\ref{fig:Oersted_creation}) of $1$. 
The same process applies to the other arcs $\mathbf{S}_{B^{\prime}}\mathbf{S}_{C^{\prime}}$,
$\mathbf{S}_{C^{\prime}}\mathbf{S}_{D^{\prime}}$, and $\mathbf{S}_{D^{\prime}}\mathbf{S}_{A^{\prime}}$.
Notice that these four arcs form a closed loop enclosing the south
pole as shown by the red curve in Fig. \ref{fig:Oersted_creation}
(c). 
Therefore, the trajectory of $\mathbf{S}_{0}$ must cross this
closed loop once, and an integer change of the topological charge
is guaranteed regardless of the actual geometry of the $\mathbf{S}_{0}$
trajectory. 
A single skyrmion is thus created. 

The critical current density, $j_{C}$, to trigger the topological
charge is shown in Fig. \ref{fig:Oersted_creation} (f). 
The value of $j_{C}$ is on the order of $10^{9}\thinspace\textrm{A}/\textrm{cm}^{2}$,
which is three orders of magnitudes larger than the typical switching
current density applicable to integrited circuits. 
An increase in
the electrode radius of several nanometers can reduce $j_{C}$, 
but it cannot provide improvements by orders of magnitudes.

\subsection{Spin transfer torque (SST) driven switching}

Since the threshold current is so high,
skyrmion creation due to a pure Oersted field is not practical.
Spin-polarization of the injected current
can reduce the threshold current density 
by one order of magnitude.
In this case the dynamical process is dominated by the STT, which
has been proposed to be a promising mechanism to switch nano-magnets
in spintronic integrited circuits \cite{behin-aein_proposal_2010}. 

\begin{figure}
\begin{centering}
\includegraphics[width=1\columnwidth]{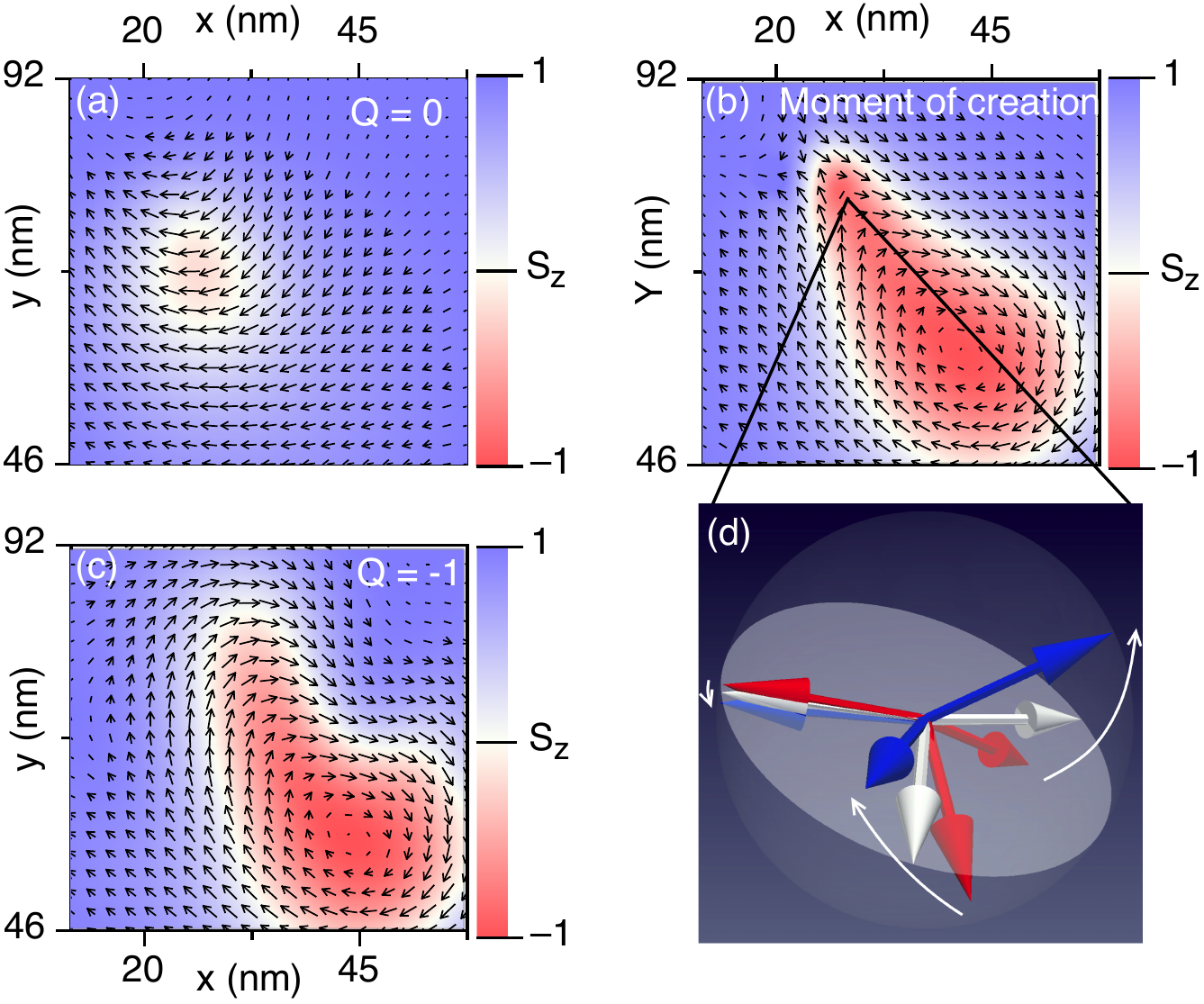} 
\par\end{centering}

\protect\protect\caption{(color online) 
(a), (b), and (c) are snap shots of the spin texture several picoseconds
around the moment of the topological transition in a Bloch-type helimagnet
thin film. (d) demonstrates the spin trajectories of the local topological
transition. The red and blue arrows denote the configurations before
and after the transition, respectively. A coplanar and non-colinear
configuration is achieved exactly at the moment of the skyrmion
creation (white arrows). 
\label{fig:DeviceSetup_SnapShots}}
\end{figure}

\begin{figure}
\begin{centering}
\includegraphics[width=1\columnwidth]{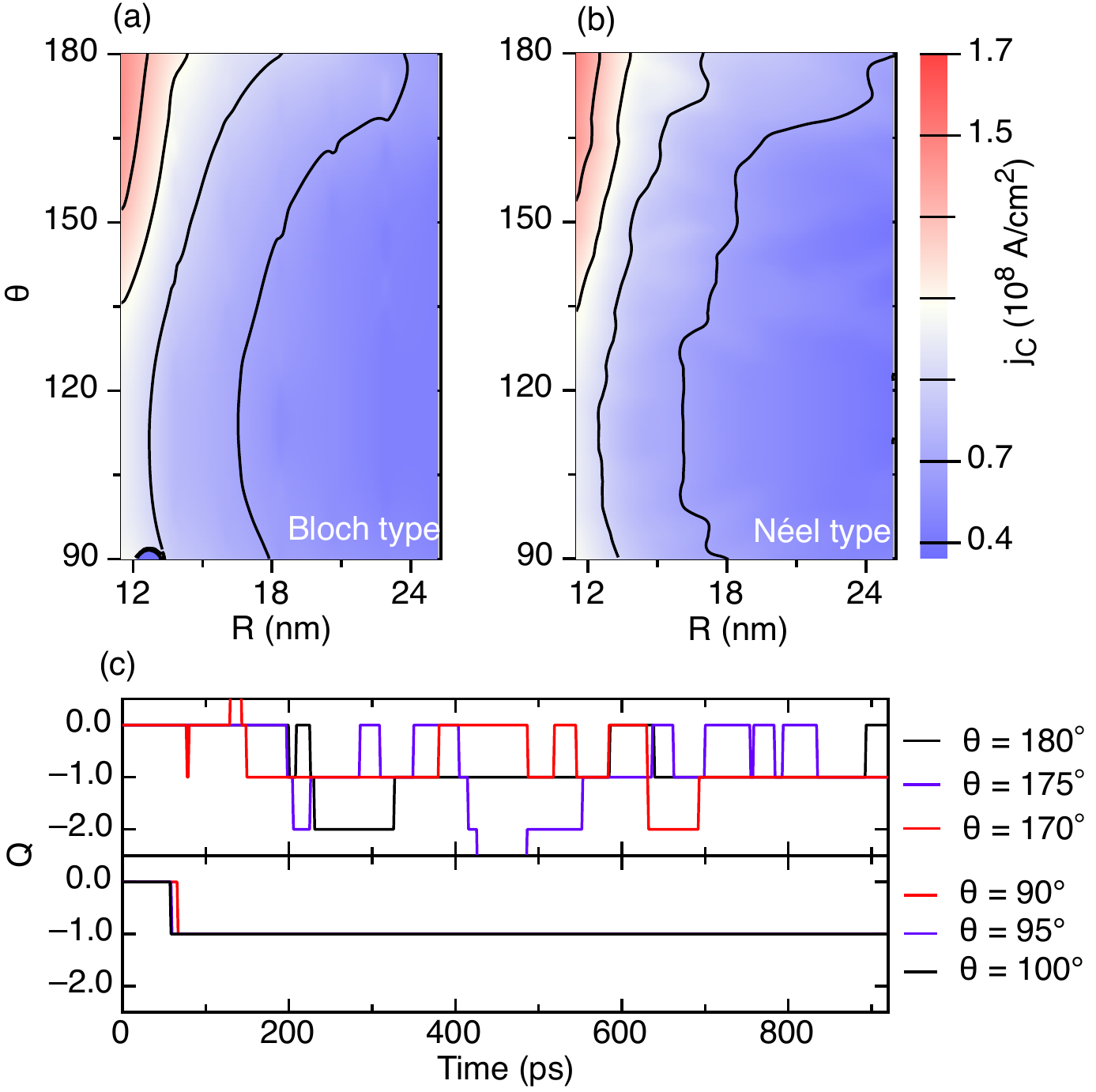} 
\par\end{centering}
\protect\protect\caption{The phase diagram of $j_{C}$ for the 
(a) Bloch-type 
and the (b) N\'{e}el-type skyrmions. 
(c) The time evolution of the topological charge at
several different polarization angles in the case of Bloch-type skyrmions. }
\label{fig:Bloch_Neel_phase_diags}
\end{figure}

Several snap shots of this dynamical process are shown in Fig.
\ref{fig:DeviceSetup_SnapShots}. 
After the current is turned on,
the STT and the Oersted field drive the spins into the x-y plane near
the electrode.
Since spins at the periphery
deviate from the outside FM configuration, the DM energy
starts to increase. 
This drives the spin texture to form a bubble-like
domain, in which the center spins present negative $z$ components,
while the spins at the periphery give large in-plane components. 
The bubble-like domain then continues to grow and starts to wrap into
a circular domain wall with a singularity. 
Around the singularity,
the spins gradually develop into an anti-parallel configuration, which
then generate a quick, drastic dynamical process that creates a topological
charge of $-1$. 
In contrast to the creation induced by pure Oersted
fields, the STT triggerred creation works with both Bloch and N\'{e}el type skyrmions. 
Movies containing the details of this dynamical process are available
in the supplimentary materials. 
The spin trajectories corresponding
to the local topological transition are shown in 
Fig. \ref{fig:DeviceSetup_SnapShots} (e), 
which follows the coplanar and non-collinear configuration discussed
in the previous section.
The critical current density to trigger the skyrmion creation event
is evaluated through a series of LLG simulations. 
For a low current density, 
excitations damp away very fast, and no skyrmion is created. 
The creation of skyrmions occurs only when the current density reaches
a critical value, $j_{C}$. 
The phase diagram of $j_{C}$ is a function
of the spin polarization angle, $\theta$, and the electrode radius,
$R$, as shown in Fig. \ref{fig:Bloch_Neel_phase_diags}. 
Here $R$ varies from $11.5\thinspace\textrm{nm}$ to $25.3\thinspace\textrm{nm}$,
while the polarization of the injected current is modifed from $-\hat{z}$
to the $x\textrm{-}y$ plane ($90^{\circ}<\theta<180^{\circ}$). 
The skyrmion creation does not happen when $\theta<90^{\circ}$. 
In this calculation, both the N\'{e}el-type and the Bloch-type skyrmion creations
are examined using the same set of parameters. 
Despite the differences
in the spin dynamical details, the phase diagrams for the two types
of skyrmions are quite similar. 
The minimum current density occurs
at $\theta\sim110^{\circ}$, where the polarization is close to the
in-plane case. 
For both skyrmion types, $j_{C}$ is approximately
$10^{8}\thinspace\textrm{A/cm}^{2}$, which is similar
to the critical switching current density due to spin orbit torques
estimated by previous numerical estimates. 
Increasing the electrode
radius can further decrease the current density but only on a linear
scale rather than an exponential scale.
The reliability and the dynamical details of the switching process
significantly depends on $\theta$, the orientation of the spin polarization. 
The STT can generate an 'anti-damping' effect during the precession
of the local magnetic moment when the injected spin is anti-parallel
to the precession axis \cite{brataas_current-induced_2012}. 
The anti-damping
can either induce a consistent oscillation or even the switching of
a single-domain nanomagnet. 
This is similar to the switching of a
single skyrmion in our proposed scheme. 
In the case of $\theta<90^{\circ}$,
the excitations induced by the torque damp away so quickly that no
topological transition could occur with a reasonable current density. 
In the case of $90^{\circ}<\theta<180^{\circ}$, switching becomes
possible. 
Since the STT is given by $\mathbf{S}\times\left(\mathbf{S}\times\boldsymbol{\sigma}\right)$,
the maximum value of the torque at $t=0$ occurs when $\theta=90^{\circ}$. 
The torque becomes zero in the case of $\theta\rightarrow180^{\circ}$,
where the required current density reaches its maximum in the phase
diagram. 
In the case of a large $\theta$, the dynamical effect continues
after the skyrmion is created, where the oscillation of the topological
charge occurs. 
The time evolution of the topological charge of a Bloch-type
skyrmion is presented in Fig. \ref{fig:EasyPlane_HeatAsist_Stability}
(b), in which several different angles are examined close to $\theta=90^{\circ}$
and $\theta=180^{\circ}$. 
Within $800\thinspace\textrm{ps}$, topological
charge (sometimes more than 1) quickly switch on and off in the case
of $\theta\sim180^{\circ}$ due to the constant oscillations driven
by the STT. 
The final state is highly sensitive to the duration of
the applied current and the details of the geometry. 
On the other
hand, the topological charge becomes stable in the case of the in-plane
polarization. 
The switching can occur in $\sim60\,\textrm{ps}$, after
which no further excitations can be initiated and no change of the
skyrmion number is witnessed. 
A similar trend is also observed in
the N\'{e}el-type skyrmions. 
This observation is consistent with the skyrmion
number oscillation observed in \cite{romming_writing_2013}. 
Small changes
in the angle do not strongly affect the switching outcome, indicating
that the in-plane polarization is best for creating skyrmions for application
purposes.

\subsection{Easy-plane anisotropy and heating effects}

\begin{figure}
\begin{centering}
\includegraphics[width=1\columnwidth]{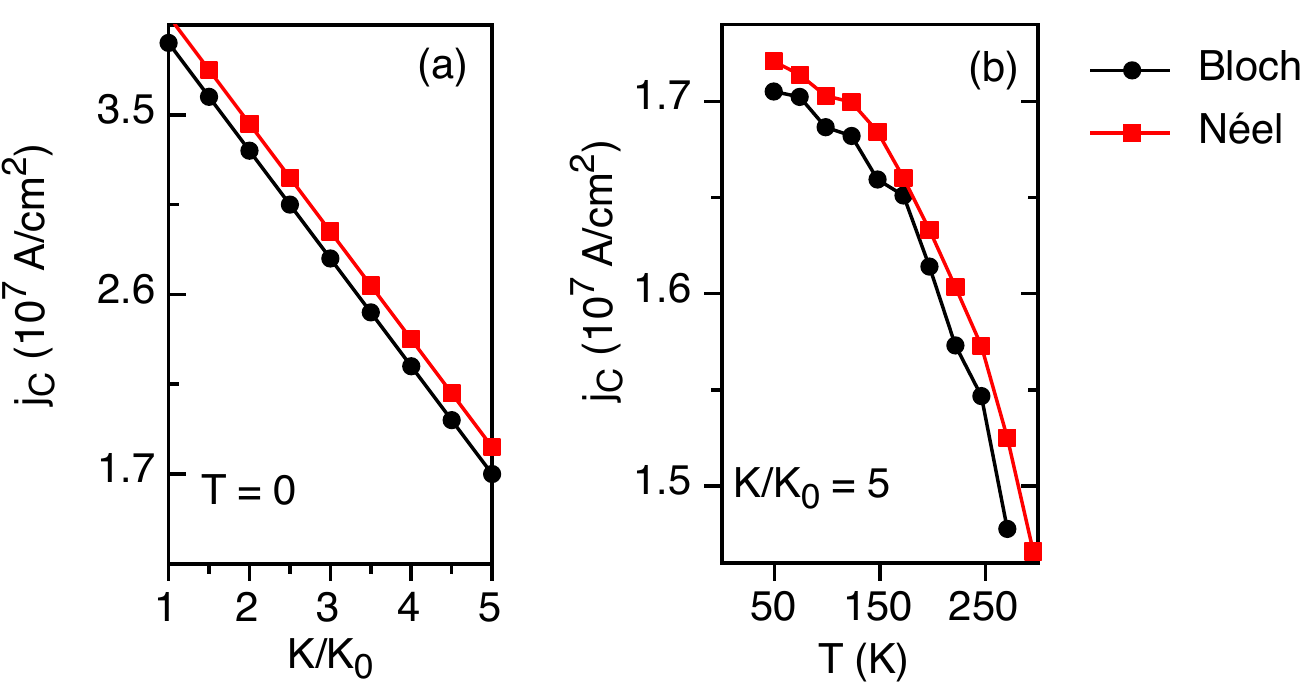} 
\par\end{centering}
\protect\protect\caption{(a) The reduction of $j_{C}$ at different values of $K$. (b) The
heat assisted skyrmion creation at finite temperatures. Each point
is an average over 400 different sampling runs. \label{fig:EasyPlane_HeatAsist_Stability}}
\end{figure}

Although the current density of $10^{8}\thinspace\textrm{A}/\textrm{cm}^{2}$
is still difficult to achieve in applications, the threshold current
can be further reduced by an order of magnitude due to easy-plane
anisotropy and heating effects. 
The easy-plane uniaxial anisotropy
is written in the Hamiltonian as $H_{\textrm{ansi}}=\sum_{i}KV\left(S_{z}^{i}\right)^{2}$
where $K$ is the anisotropy energy density and $V$ is the volume
of each local spin. 
This term is physically induced by a combination
of the strained structrual effects at the interface and the demagnetization
effects due to the aspect ratio. 
In a helimagnet, it has been proposed
that $K$ is measured by $K_{0}$, the effective stiffness of the
conical phase determined by material parameters (for FeGe, $K_{0}=1.7\times10^{3}\thinspace\textrm{J/m}^{3}$)
\cite{karhu_chiral_2012}. 
Recent experimental results indicate that
the skyrmion phase in a FeGe thin film can be significantly extended
in the phase diagram, and the value of $\nicefrac{K}{K_{0}}$ reaches
$\thicksim1$ when the thickness reduces to $5\thinspace\textrm{nm}$
\cite{huang_extended_2012}. 
Larger values of anisotropy are expected
if the thickness further decreases. 
Since the anisotropy energetically
prefers the in-plane configuration, it helps the spin transfer torque
to drive the spins to reach the coplanar switching configuration. 
The required current density can thus be reduced. Starting from the
optimum situation in the phase diagram ($\theta=110^{\circ}$, $R=25.3\thinspace\textrm{nm}$),
the value of $\nicefrac{K}{K_{0}}$ is modifed from $1$ to $5$ in
our calculation. 
As shown in Fig. \ref{fig:EasyPlane_HeatAsist_Stability} (a), 
the easy-plane anisotropy reduces the switching current density
from by approximately a factor of 2.2 for both types of skyrmions. 

The thermal fluctuation given by a finite temperature is further examined
numerically. 
In order to include this effect, a stochastic field $\mathbf{L}$
is added onto the effective field in Eq. (\ref{eq:LLG}) 
\cite{garcia-palacios_langevin-dynamics_1998}. 
The dissipation-fluctuation relation $\left\langle L_{\mu}(\mathbf{r},t)L_{\nu}(\mathbf{r}^{\prime},t^{\prime})\right\rangle =\xi\delta_{\mu\nu}\delta_{\mathbf{rr}^{\prime}}\delta_{tt^{\prime}}$
is satisfied, where $\xi=\alpha k_{\text{B}}T/\gamma$, and $T$ is
the temperature. 
The average $\left\langle \thinspace\right\rangle $
is taken over the realizations of the fluctuation field. 
The deterministic
Heun scheme is employed to integrate out this stochastic LLG equation. 
Below $T_{C}$ ($270\thinspace\textrm{K}$ in FeGe), the average switching
current density based on $400$ sampling runs is obtained for both
the N\'{e}el-type and the Bloch-type skyrmions. 
The results are shown
in Fig. \ref{fig:EasyPlane_HeatAsist_Stability}(c). 
 Movies recording
this process can be found in the supplementary material. 
Although
thermal fluctuations randomize the local spins at each time step,
the overall dynamical process of the skyrmion creation is similar
to that at the zero temperature. 
This demonstrates the stability of
the skyrmion creation scheme below $T_{C}$. 
The average switching
current density decreases slightly due to the thermal fluctuations,
indicating a negligible heat asisted switching effect. 
Above $T_{C}$,
although the switching current can be further reduced, the magnetic
order starts to vanish, where random topological charges can be spontaneously
excited by thermal fluctuations. 
This should be avoided in the proposed switching scheme.

\section{topological protection}

\begin{figure}
\begin{centering}
\includegraphics[width=1\columnwidth]{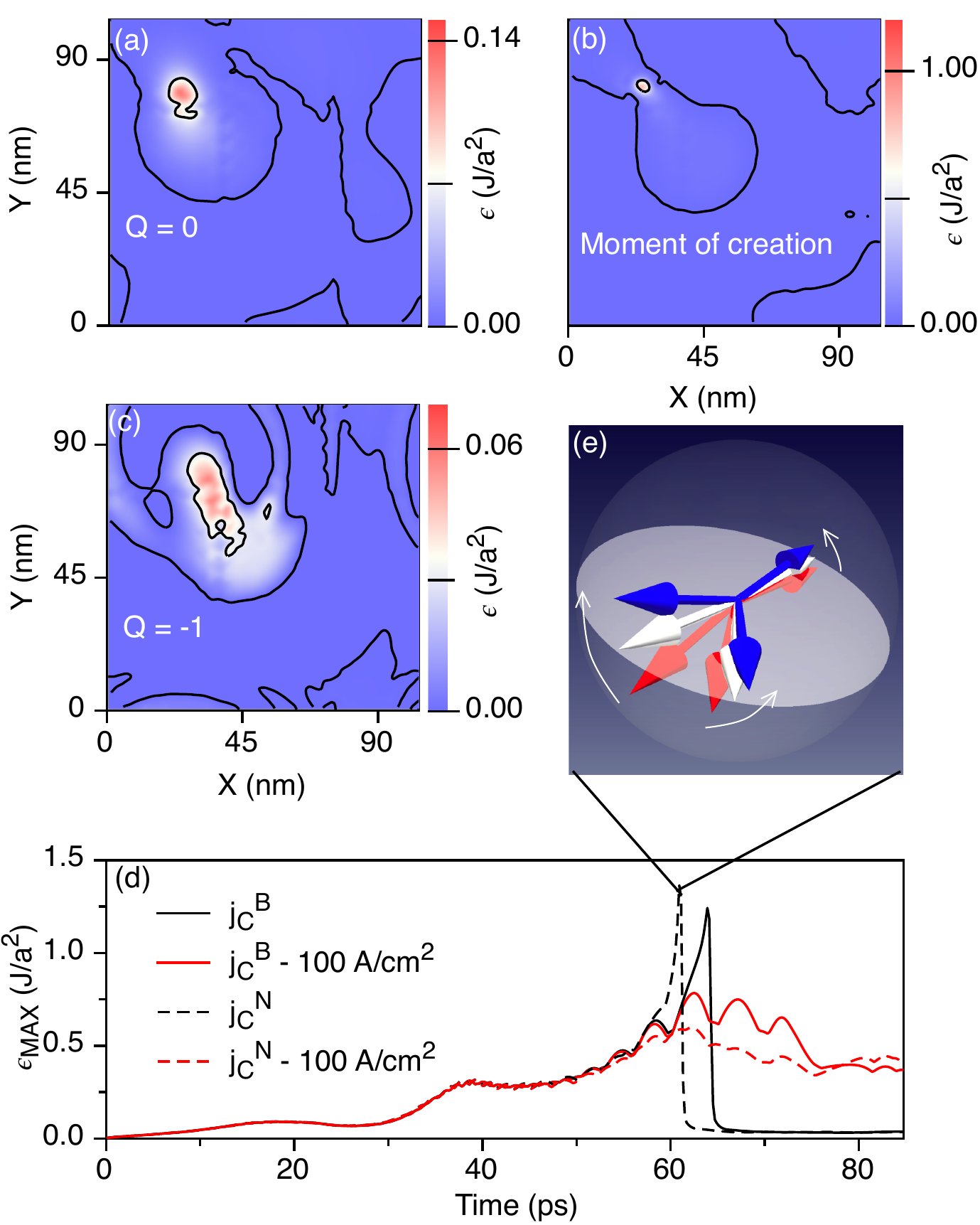} 
\par\end{centering}
\protect\protect\caption{
The energy landscape of the topological transition. (a), (b) and (c)
illustrate the energy density distribution of a Bloch-type skyrmion
creation process. 
These plots correspond to the snap shots given in
Fig. \ref{fig:DeviceSetup_SnapShots} (b), (c) and (d). 
The time evolution of the maximum energy density for both types
of skyrmions is plotted in (d). 
(e) illustrates the critical spin configuration when a N\'{e}el-type
skyrmion is created. 
\label{fig:EnergyLandscape}}
\end{figure}

The difficulty of skyrmion switching originates from the critical
spin configuration required by the topological transition. 
The energy landscapes
several picoseconds around the moment of creation are shown
in Fig. \ref{fig:EnergyLandscape} (a)-(c). 
Exactly at the transition moment, energy is highly concentrated at
the switching position, where the energy density overcomes the minimum
topological energy barrier, $J$. 
For both N\'{e}el-type and the Bloch-type
skyrmion switching, the maxmium energy density evolves through time,
which is plotted in Fig. \ref{fig:EnergyLandscape}(d). 
Even when
the injected current density is only 100\,$\textrm{A}/\textrm{cm}^{2}$
below $j_{C}$, the energy density cannot overcome the topological
barrier, and no skyrmion can be created. 
Both of the two cases
present similar line shapes of the energy evolution, despite the significant
differences in the dynamical details (see the movies in the supplementary
material). 
In both cases, fast skyrmion switching within $\sim60\thinspace\textrm{ps}$
is achieved. 
The barrier height of the N\'{e}el type is a little larger
than that of the Bloch type, which is determined by the exact switching
configuration. 
As shown in Fig. \ref{fig:EnergyLandscape}(e), the
spin alignment is more non-colinear compared to that given in Fig.
\ref{fig:DeviceSetup_SnapShots}(e), contributing more exchange energy
than that of the Bloch type. 
The difference in this configuration
comes from the swirling Oersted field induced by the vertical current. 
For the Bloch-type skyrmions, the Oersted field helps the in-plane
DM interaction form the co-planar texture, while it does not assist
the out-of-plane DM interaction in the N\'{e}el-type skyrmions. 
As shown in the phase diagrams in Fig. \ref{fig:Bloch_Neel_phase_diags}, 
this difference in the barrier does not significantly
affect the switching current density. 
\begin{figure}
\begin{centering}
\includegraphics[width=1\columnwidth]{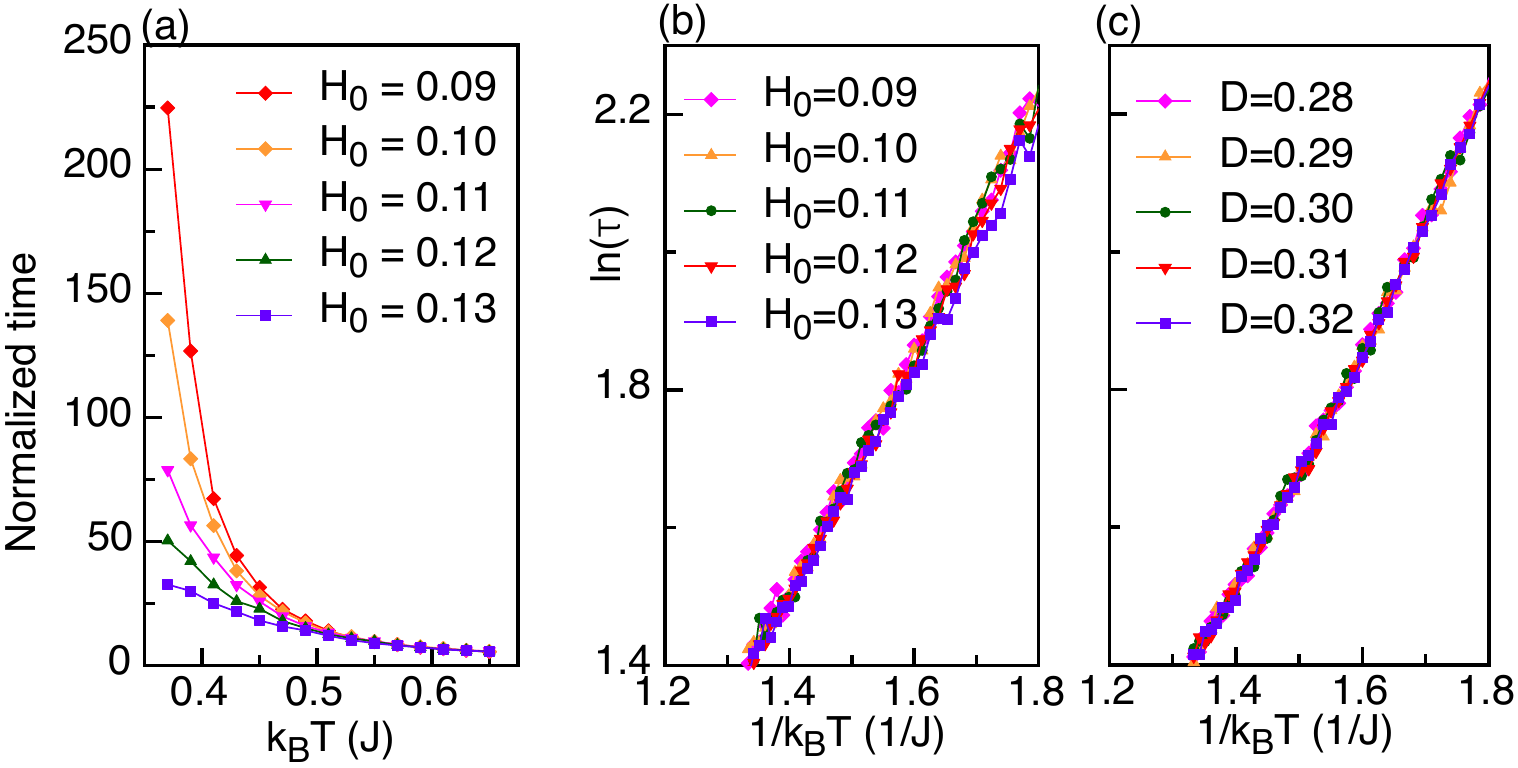} 
\par\end{centering}
\protect\protect\caption{Average life time of a topological charge induced by thermal fluctuations.
(a), Single Skyrmion life time at finite temperature ($D=0.3J$, $\alpha=0.1)$.
Each point is an average of $1000$ sampling runs. (b)-(c), $\ln(\tau)$
vs. $1/k_{B}T$ at different $H_{0}$ and $D$. The slope of the linear
fit gives a numerical estimation of $E_{a}\sim1.7J$.
\label{fig:Average-life-time}}
\end{figure}

This estimate of the topological protection barrier can also be numerically
determined from the thermal activation energy. 
The activation energy of each
topological charge is extracted numerically by examining the lifetime
of a topological charge as a function of temperature. 
A skyrmion is
simulated at finite temperature until the topological charge switches
from -1 to 0 due to the random thermal fluctuations. 
The time of the
annihilation is recorded. 
This simulation is repeated 1000 times at
each temperature, and the average lifetime $\tau$ is determined as
a function of temperature. 
Plots of $\tau$ versus temperature for
different background fields $H_0$ are shown in Fig. \ref{fig:Average-life-time}
(a). 
At low temperatures, a smaller $H_{0}$ results in a more stable
single skyrmion with a longer lifetime. 
At higher temperatures such
that $k_{B}T$ approaches $J$, all of the curves in Fig. \ref{fig:Average-life-time}
(a) converge and decay exponentially. 
For the transition from skyrmion
to ferromagnet, the transition rate $k$ obeys the Arrhenius equation
$k\sim\exp(-E_{\text{a}}/k_{B}T)$. 
The lifetime $\tau$ is the inverse
of $k$ so that $\tau\sim1/k\sim\exp(E_{\text{a}}/k_{B}T)$. 
The plot
of $\ln\tau$ in Fig. \ref{fig:Average-life-time} (b) is linear in
the inverse temperature $1/T$. Plots for various $H_{0}$ are nearly
parallel with each other with an activation energy of $E_{\text{a}}\sim1.7J$. 
Further analysis shows that this activation energy is also insensitive
to the DM interaction as shown in Fig. \ref{fig:Average-life-time}
(c). 
The activation energy $E_{\text{a}}\sim1.7J$ determined from the Arrhenius
plots represents the barrier to decay in a skyrmion annihilation process.
The maximum activation energy density of 1.25 $J$ per spin determined
by direct calculation in Fig. \ref{fig:EnergyLandscape}(d) 
represents the energy barrier to creation.
These values fall within the range $J<\Delta\epsilon<2J$
determined from the topological charge analysis leading to
Eq. (\ref{eq:Barrier}),
and they support the picture of the topological origin of the activation
energy that stabilizes the single skyrmion.

\section{Conclusions}

A topological charge analysis provides insight into
the locally triggered transition from a trivial to a non-trivial topological
spin texture of the N\'{e}el or Bloch type skyrmion.
The topological protection 
of the magnetic skyrmion 
is determined by the symmetric Heisenberg exchange energy. 
The topological charge analysis, direct micromagnetic calculation, and 
extraction from Arrhenius plots created from ensemble averaged finite
temperature calculations all give consistent values for the
energy barrier determined by the spin geometry at the point of transition
between a trivial and non-trivial spin topology of $J<\Delta\epsilon<2J$.
A scheme to create single skyrmions is analyzed for
both N\'{e}el-type and Bloch-type in helimagnetic
thin films utilizing the dynamical excitations induced by the Oersted
field and the STT given by a vertically injected spin-polarized current. 
The critical switching current density is $\sim10^{7}\thinspace\textrm{A/cm}^{2}$,
which decreases with the easy-plane type uniaxial anisotropy and
thermal fluctuations. 
In-plane spin polarization of the injected current performs better than 
out-of-plane polarization, and it
provides ultrafast switching times (within $100\thinspace\textrm{ps}$)
and reliable switching outcomes. 
\begin{acknowledgments}
The micromagnetic simulations, analysis, and writing were supported by the National Science Foundation Grant No. 1408168. The analysis and writing were also supported in part by FAME, one of six centers of STARnet, a Semiconductor Research Corporation program sponsored by MARCO and DARPA. The device design and the experimental discussion were supported in part by the  Spins and Heat in Nanoscale Electronic Systems (Spins) an Energy Frontier Research Center funded by the U.S. Department of Energy, Office of Science, Basic Energy Sciences under Award No. DE-SC0012670.
\end{acknowledgments}


\begin{thebibliography}{32}
\expandafter\ifx\csname natexlab\endcsname\relax\def\natexlab#1{#1}\fi
\expandafter\ifx\csname bibnamefont\endcsname\relax
  \def\bibnamefont#1{#1}\fi
\expandafter\ifx\csname bibfnamefont\endcsname\relax
  \def\bibfnamefont#1{#1}\fi
\expandafter\ifx\csname citenamefont\endcsname\relax
  \def\citenamefont#1{#1}\fi
\expandafter\ifx\csname url\endcsname\relax
  \def\url#1{\texttt{#1}}\fi
\expandafter\ifx\csname urlprefix\endcsname\relax\def\urlprefix{URL }\fi
\providecommand{\bibinfo}[2]{#2}
\providecommand{\eprint}[2][]{\url{#2}}

\bibitem[{\citenamefont{Nagaosa and Tokura}(2013)}]{Nagaosa_NatNano_Review}
\bibinfo{author}{\bibfnamefont{N.}~\bibnamefont{Nagaosa}} \bibnamefont{and}
  \bibinfo{author}{\bibfnamefont{Y.}~\bibnamefont{Tokura}},
  \bibinfo{journal}{Nat Nano} \textbf{\bibinfo{volume}{8}},
  \bibinfo{pages}{899} (\bibinfo{year}{2013}).

\bibitem[{\citenamefont{R{\"o}{\ss}ler
  et~al.}(2006)\citenamefont{R{\"o}{\ss}ler, Bogdanov, and
  Pfleiderer}}]{rosler_spontaneous_2006}
\bibinfo{author}{\bibfnamefont{U.~K.} \bibnamefont{R{\"o}{\ss}ler}},
  \bibinfo{author}{\bibfnamefont{A.~N.} \bibnamefont{Bogdanov}},
  \bibnamefont{and}
  \bibinfo{author}{\bibfnamefont{C.}~\bibnamefont{Pfleiderer}},
  \bibinfo{journal}{Nature} \textbf{\bibinfo{volume}{442}},
  \bibinfo{pages}{797} (\bibinfo{year}{2006}), ISSN \bibinfo{issn}{0028-0836}.

\bibitem[{\citenamefont{M{\"u}hlbauer et~al.}(2009)\citenamefont{M{\"u}hlbauer,
  Binz, Jonietz, Pfleiderer, Rosch, Neubauer, Georgii, and
  B{\"o}ni}}]{muhlbauer_skyrmion_2009}
\bibinfo{author}{\bibfnamefont{S.}~\bibnamefont{M{\"u}hlbauer}},
  \bibinfo{author}{\bibfnamefont{B.}~\bibnamefont{Binz}},
  \bibinfo{author}{\bibfnamefont{F.}~\bibnamefont{Jonietz}},
  \bibinfo{author}{\bibfnamefont{C.}~\bibnamefont{Pfleiderer}},
  \bibinfo{author}{\bibfnamefont{A.}~\bibnamefont{Rosch}},
  \bibinfo{author}{\bibfnamefont{A.}~\bibnamefont{Neubauer}},
  \bibinfo{author}{\bibfnamefont{R.}~\bibnamefont{Georgii}}, \bibnamefont{and}
  \bibinfo{author}{\bibfnamefont{P.}~\bibnamefont{B{\"o}ni}},
  \bibinfo{journal}{Science} \textbf{\bibinfo{volume}{323}},
  \bibinfo{pages}{915} (\bibinfo{year}{2009}), ISSN \bibinfo{issn}{0036-8075,
  1095-9203}, \bibinfo{note}{{PMID:} 19213914}.

\bibitem[{\citenamefont{Yu et~al.}(2010)\citenamefont{Yu, Onose, Kanazawa,
  Park, Han, Matsui, Nagaosa, and Tokura}}]{yu_real-space_2010}
\bibinfo{author}{\bibfnamefont{X.~Z.} \bibnamefont{Yu}},
  \bibinfo{author}{\bibfnamefont{Y.}~\bibnamefont{Onose}},
  \bibinfo{author}{\bibfnamefont{N.}~\bibnamefont{Kanazawa}},
  \bibinfo{author}{\bibfnamefont{J.~H.} \bibnamefont{Park}},
  \bibinfo{author}{\bibfnamefont{J.~H.} \bibnamefont{Han}},
  \bibinfo{author}{\bibfnamefont{Y.}~\bibnamefont{Matsui}},
  \bibinfo{author}{\bibfnamefont{N.}~\bibnamefont{Nagaosa}}, \bibnamefont{and}
  \bibinfo{author}{\bibfnamefont{Y.}~\bibnamefont{Tokura}},
  \bibinfo{journal}{Nature} \textbf{\bibinfo{volume}{465}},
  \bibinfo{pages}{901} (\bibinfo{year}{2010}), ISSN \bibinfo{issn}{0028-0836}.

\bibitem[{\citenamefont{Yu et~al.}(2011)\citenamefont{Yu, anazawa, Onose,
  Kimoto, Zhang, Ishiwata, Matsui, and Tokura}}]{yu_near_2011}
\bibinfo{author}{\bibfnamefont{X.~Z.} \bibnamefont{Yu}},
  \bibinfo{author}{\bibfnamefont{N.~K.} \bibnamefont{anazawa}},
  \bibinfo{author}{\bibfnamefont{Y.}~\bibnamefont{Onose}},
  \bibinfo{author}{\bibfnamefont{K.}~\bibnamefont{Kimoto}},
  \bibinfo{author}{\bibfnamefont{W.~Z.} \bibnamefont{Zhang}},
  \bibinfo{author}{\bibfnamefont{S.}~\bibnamefont{Ishiwata}},
  \bibinfo{author}{\bibfnamefont{Y.}~\bibnamefont{Matsui}}, \bibnamefont{and}
  \bibinfo{author}{\bibfnamefont{Y.}~\bibnamefont{Tokura}},
  \bibinfo{journal}{Nat Mater} \textbf{\bibinfo{volume}{10}},
  \bibinfo{pages}{106 } (\bibinfo{year}{2011}).

\bibitem[{\citenamefont{Dzyaloshinsky}(1958)}]{dzyaloshinsky_thermodynamic_1958}
\bibinfo{author}{\bibfnamefont{I.}~\bibnamefont{Dzyaloshinsky}},
  \bibinfo{journal}{Journal of Physics and Chemistry of Solids}
  \textbf{\bibinfo{volume}{4}}, \bibinfo{pages}{241} (\bibinfo{year}{1958}),
  ISSN \bibinfo{issn}{0022-3697}.

\bibitem[{\citenamefont{Moriya}(1960)}]{moriya_anisotropic_1960}
\bibinfo{author}{\bibfnamefont{T.}~\bibnamefont{Moriya}},
  \bibinfo{journal}{Physical Review} \textbf{\bibinfo{volume}{120}},
  \bibinfo{pages}{91{\textendash}98} (\bibinfo{year}{1960}).

\bibitem[{\citenamefont{Kanazawa et~al.}(2011)\citenamefont{Kanazawa, Onose,
  Arima, Okuyama, Ohoyama, Wakimoto, Kakurai, Ishiwata, and
  Tokura}}]{kanazawa_large_2011-1}
\bibinfo{author}{\bibfnamefont{N.}~\bibnamefont{Kanazawa}},
  \bibinfo{author}{\bibfnamefont{Y.}~\bibnamefont{Onose}},
  \bibinfo{author}{\bibfnamefont{T.}~\bibnamefont{Arima}},
  \bibinfo{author}{\bibfnamefont{D.}~\bibnamefont{Okuyama}},
  \bibinfo{author}{\bibfnamefont{K.}~\bibnamefont{Ohoyama}},
  \bibinfo{author}{\bibfnamefont{S.}~\bibnamefont{Wakimoto}},
  \bibinfo{author}{\bibfnamefont{K.}~\bibnamefont{Kakurai}},
  \bibinfo{author}{\bibfnamefont{S.}~\bibnamefont{Ishiwata}}, \bibnamefont{and}
  \bibinfo{author}{\bibfnamefont{Y.}~\bibnamefont{Tokura}},
  \bibinfo{journal}{Physical Review Letters} \textbf{\bibinfo{volume}{106}},
  \bibinfo{pages}{156603} (\bibinfo{year}{2011}).

\bibitem[{\citenamefont{Han et~al.}(2010)\citenamefont{Han, Zang, Yang, Park,
  and Nagaosa}}]{han_skyrmion_2010}
\bibinfo{author}{\bibfnamefont{J.~H.} \bibnamefont{Han}},
  \bibinfo{author}{\bibfnamefont{J.}~\bibnamefont{Zang}},
  \bibinfo{author}{\bibfnamefont{Z.}~\bibnamefont{Yang}},
  \bibinfo{author}{\bibfnamefont{J.-H.} \bibnamefont{Park}}, \bibnamefont{and}
  \bibinfo{author}{\bibfnamefont{N.}~\bibnamefont{Nagaosa}},
  \bibinfo{journal}{Physical Review B} \textbf{\bibinfo{volume}{82}},
  \bibinfo{pages}{094429} (\bibinfo{year}{2010}).

\bibitem[{\citenamefont{Munzer et~al.}(2010)\citenamefont{Munzer, Neubauer,
  Adams, Muhlbauer, Franz, Jonietz, Georgii, Boni, Pedersen, Schmidt
  et~al.}}]{munzer_skyrmion_2010}
\bibinfo{author}{\bibfnamefont{W.}~\bibnamefont{Munzer}},
  \bibinfo{author}{\bibfnamefont{A.}~\bibnamefont{Neubauer}},
  \bibinfo{author}{\bibfnamefont{T.}~\bibnamefont{Adams}},
  \bibinfo{author}{\bibfnamefont{S.}~\bibnamefont{Muhlbauer}},
  \bibinfo{author}{\bibfnamefont{C.}~\bibnamefont{Franz}},
  \bibinfo{author}{\bibfnamefont{F.}~\bibnamefont{Jonietz}},
  \bibinfo{author}{\bibfnamefont{R.}~\bibnamefont{Georgii}},
  \bibinfo{author}{\bibfnamefont{P.}~\bibnamefont{Boni}},
  \bibinfo{author}{\bibfnamefont{B.}~\bibnamefont{Pedersen}},
  \bibinfo{author}{\bibfnamefont{M.}~\bibnamefont{Schmidt}},
  \bibnamefont{et~al.}, \bibinfo{journal}{Phys. Rev. B}
  \textbf{\bibinfo{volume}{81}}, \bibinfo{pages}{041203}
  (\bibinfo{year}{2010}).

\bibitem[{\citenamefont{Seki et~al.}(2012)\citenamefont{Seki, Yu, Ishiwata, and
  Tokura}}]{seki_observation_2012}
\bibinfo{author}{\bibfnamefont{S.}~\bibnamefont{Seki}},
  \bibinfo{author}{\bibfnamefont{X.~Z.} \bibnamefont{Yu}},
  \bibinfo{author}{\bibfnamefont{S.}~\bibnamefont{Ishiwata}}, \bibnamefont{and}
  \bibinfo{author}{\bibfnamefont{Y.}~\bibnamefont{Tokura}},
  \bibinfo{journal}{Science} \textbf{\bibinfo{volume}{336}},
  \bibinfo{pages}{198 } (\bibinfo{year}{2012}).

\bibitem[{\citenamefont{Yu et~al.}(2012)\citenamefont{Yu, Kanazawa, Zhang,
  Nagai, Hara, Kimoto, Matsui, Onose, and Tokura}}]{yu_skyrmion_2012}
\bibinfo{author}{\bibfnamefont{X.~Z.} \bibnamefont{Yu}},
  \bibinfo{author}{\bibfnamefont{N.}~\bibnamefont{Kanazawa}},
  \bibinfo{author}{\bibfnamefont{W.~Z.} \bibnamefont{Zhang}},
  \bibinfo{author}{\bibfnamefont{T.}~\bibnamefont{Nagai}},
  \bibinfo{author}{\bibfnamefont{T.}~\bibnamefont{Hara}},
  \bibinfo{author}{\bibfnamefont{K.}~\bibnamefont{Kimoto}},
  \bibinfo{author}{\bibfnamefont{Y.}~\bibnamefont{Matsui}},
  \bibinfo{author}{\bibfnamefont{Y.}~\bibnamefont{Onose}}, \bibnamefont{and}
  \bibinfo{author}{\bibfnamefont{Y.}~\bibnamefont{Tokura}},
  \bibinfo{journal}{Nature Communications} \textbf{\bibinfo{volume}{3}},
  \bibinfo{pages}{988} (\bibinfo{year}{2012}).

\bibitem[{\citenamefont{Li et~al.}(2013)\citenamefont{Li, Kanazawa, Yu,
  Tsukazaki, Kawasaki, Ichikawa, Jin, Kagawa, and Tokura}}]{li_robust_2013}
\bibinfo{author}{\bibfnamefont{Y.}~\bibnamefont{Li}},
  \bibinfo{author}{\bibfnamefont{N.}~\bibnamefont{Kanazawa}},
  \bibinfo{author}{\bibfnamefont{X.~Z.} \bibnamefont{Yu}},
  \bibinfo{author}{\bibfnamefont{A.}~\bibnamefont{Tsukazaki}},
  \bibinfo{author}{\bibfnamefont{M.}~\bibnamefont{Kawasaki}},
  \bibinfo{author}{\bibfnamefont{M.}~\bibnamefont{Ichikawa}},
  \bibinfo{author}{\bibfnamefont{X.~F.} \bibnamefont{Jin}},
  \bibinfo{author}{\bibfnamefont{F.}~\bibnamefont{Kagawa}}, \bibnamefont{and}
  \bibinfo{author}{\bibfnamefont{Y.}~\bibnamefont{Tokura}},
  \bibinfo{journal}{Physical Review Letters} \textbf{\bibinfo{volume}{110}},
  \bibinfo{pages}{117202} (\bibinfo{year}{2013}).

\bibitem[{\citenamefont{Heinze et~al.}(2011)\citenamefont{Heinze, von Bergmann,
  Menzel, Brede, Kubetzka, Wiesendanger, Bihlmayer, and
  Bl{\"u}gel}}]{heinze_spontaneous_2011}
\bibinfo{author}{\bibfnamefont{S.}~\bibnamefont{Heinze}},
  \bibinfo{author}{\bibfnamefont{K.}~\bibnamefont{von Bergmann}},
  \bibinfo{author}{\bibfnamefont{M.}~\bibnamefont{Menzel}},
  \bibinfo{author}{\bibfnamefont{J.}~\bibnamefont{Brede}},
  \bibinfo{author}{\bibfnamefont{A.}~\bibnamefont{Kubetzka}},
  \bibinfo{author}{\bibfnamefont{R.}~\bibnamefont{Wiesendanger}},
  \bibinfo{author}{\bibfnamefont{G.}~\bibnamefont{Bihlmayer}},
  \bibnamefont{and}
  \bibinfo{author}{\bibfnamefont{S.}~\bibnamefont{Bl{\"u}gel}},
  \bibinfo{journal}{Nature Physics} \textbf{\bibinfo{volume}{7}},
  \bibinfo{pages}{713} (\bibinfo{year}{2011}), ISSN \bibinfo{issn}{1745-2473}.

\bibitem[{\citenamefont{Jonietz et~al.}(2010)\citenamefont{Jonietz,
  M{\"u}hlbauer, Pfleiderer, Neubauer, M{\"u}nzer, Bauer, Adams, Georgii,
  B{\"o}ni, Duine et~al.}}]{jonietz_spin_2010}
\bibinfo{author}{\bibfnamefont{F.}~\bibnamefont{Jonietz}},
  \bibinfo{author}{\bibfnamefont{S.}~\bibnamefont{M{\"u}hlbauer}},
  \bibinfo{author}{\bibfnamefont{C.}~\bibnamefont{Pfleiderer}},
  \bibinfo{author}{\bibfnamefont{A.}~\bibnamefont{Neubauer}},
  \bibinfo{author}{\bibfnamefont{W.}~\bibnamefont{M{\"u}nzer}},
  \bibinfo{author}{\bibfnamefont{A.}~\bibnamefont{Bauer}},
  \bibinfo{author}{\bibfnamefont{T.}~\bibnamefont{Adams}},
  \bibinfo{author}{\bibfnamefont{R.}~\bibnamefont{Georgii}},
  \bibinfo{author}{\bibfnamefont{P.}~\bibnamefont{B{\"o}ni}},
  \bibinfo{author}{\bibfnamefont{R.~A.} \bibnamefont{Duine}},
  \bibnamefont{et~al.}, \bibinfo{journal}{Science}
  \textbf{\bibinfo{volume}{330}}, \bibinfo{pages}{1648} (\bibinfo{year}{2010}),
  ISSN \bibinfo{issn}{0036-8075, 1095-9203}, \bibinfo{note}{{PMID:} 21164010}.

\bibitem[{\citenamefont{Schulz et~al.}(2012)\citenamefont{Schulz, Ritz, Bauer,
  Halder, Wagner, Franz, Pfleiderer, Everschor, Garst, and
  Rosch}}]{schulz_emergent_2012}
\bibinfo{author}{\bibfnamefont{T.}~\bibnamefont{Schulz}},
  \bibinfo{author}{\bibfnamefont{R.}~\bibnamefont{Ritz}},
  \bibinfo{author}{\bibfnamefont{A.}~\bibnamefont{Bauer}},
  \bibinfo{author}{\bibfnamefont{M.}~\bibnamefont{Halder}},
  \bibinfo{author}{\bibfnamefont{M.}~\bibnamefont{Wagner}},
  \bibinfo{author}{\bibfnamefont{C.}~\bibnamefont{Franz}},
  \bibinfo{author}{\bibfnamefont{C.}~\bibnamefont{Pfleiderer}},
  \bibinfo{author}{\bibfnamefont{K.}~\bibnamefont{Everschor}},
  \bibinfo{author}{\bibfnamefont{M.}~\bibnamefont{Garst}}, \bibnamefont{and}
  \bibinfo{author}{\bibfnamefont{A.}~\bibnamefont{Rosch}},
  \bibinfo{journal}{Nature Physics} \textbf{\bibinfo{volume}{8}},
  \bibinfo{pages}{301} (\bibinfo{year}{2012}), ISSN \bibinfo{issn}{1745-2473}.

\bibitem[{\citenamefont{Zang et~al.}(2011)\citenamefont{Zang, Mostovoy, Han,
  and Nagaosa}}]{zang_dynamics_2011}
\bibinfo{author}{\bibfnamefont{J.}~\bibnamefont{Zang}},
  \bibinfo{author}{\bibfnamefont{M.}~\bibnamefont{Mostovoy}},
  \bibinfo{author}{\bibfnamefont{J.~H.} \bibnamefont{Han}}, \bibnamefont{and}
  \bibinfo{author}{\bibfnamefont{N.}~\bibnamefont{Nagaosa}},
  \bibinfo{journal}{Physical Review Letters} \textbf{\bibinfo{volume}{107}},
  \bibinfo{pages}{136804} (\bibinfo{year}{2011}).

\bibitem[{\citenamefont{Fert et~al.}(2013)\citenamefont{Fert, Cros, and
  Sampaio}}]{fert_skyrmions_2013}
\bibinfo{author}{\bibfnamefont{A.}~\bibnamefont{Fert}},
  \bibinfo{author}{\bibfnamefont{V.}~\bibnamefont{Cros}}, \bibnamefont{and}
  \bibinfo{author}{\bibfnamefont{J.}~\bibnamefont{Sampaio}},
  \bibinfo{journal}{Nature nanotechnology} \textbf{\bibinfo{volume}{8}},
  \bibinfo{pages}{152} (\bibinfo{year}{2013}).

\bibitem[{\citenamefont{Sampaio et~al.}(2013)\citenamefont{Sampaio, Cros,
  Rohart, Thiaville, and Fert}}]{sampaio_nucleation_2013}
\bibinfo{author}{\bibfnamefont{J.}~\bibnamefont{Sampaio}},
  \bibinfo{author}{\bibfnamefont{V.}~\bibnamefont{Cros}},
  \bibinfo{author}{\bibfnamefont{S.}~\bibnamefont{Rohart}},
  \bibinfo{author}{\bibfnamefont{A.}~\bibnamefont{Thiaville}},
  \bibnamefont{and} \bibinfo{author}{\bibfnamefont{A.}~\bibnamefont{Fert}},
  \bibinfo{journal}{Nature Nanotechnology} \textbf{\bibinfo{volume}{8}},
  \bibinfo{pages}{839} (\bibinfo{year}{2013}), ISSN \bibinfo{issn}{1748-3395}.

\bibitem[{\citenamefont{Iwasaki et~al.}(2013)\citenamefont{Iwasaki, Mochizuki,
  and Nagaosa}}]{iwasaki_current-induced_2013}
\bibinfo{author}{\bibfnamefont{J.}~\bibnamefont{Iwasaki}},
  \bibinfo{author}{\bibfnamefont{M.}~\bibnamefont{Mochizuki}},
  \bibnamefont{and} \bibinfo{author}{\bibfnamefont{N.}~\bibnamefont{Nagaosa}},
  \bibinfo{journal}{Nature Nanotechnology} \textbf{\bibinfo{volume}{8}},
  \bibinfo{pages}{742} (\bibinfo{year}{2013}), ISSN \bibinfo{issn}{1748-3387}.

\bibitem[{\citenamefont{Romming et~al.}(2013)\citenamefont{Romming, Hanneken,
  Menzel, Bickel, Wolter, Bergmann, Kubetzka, and
  Wiesendanger}}]{romming_writing_2013}
\bibinfo{author}{\bibfnamefont{N.}~\bibnamefont{Romming}},
  \bibinfo{author}{\bibfnamefont{C.}~\bibnamefont{Hanneken}},
  \bibinfo{author}{\bibfnamefont{M.}~\bibnamefont{Menzel}},
  \bibinfo{author}{\bibfnamefont{J.~E.} \bibnamefont{Bickel}},
  \bibinfo{author}{\bibfnamefont{B.}~\bibnamefont{Wolter}},
  \bibinfo{author}{\bibfnamefont{K.~v.} \bibnamefont{Bergmann}},
  \bibinfo{author}{\bibfnamefont{A.}~\bibnamefont{Kubetzka}}, \bibnamefont{and}
  \bibinfo{author}{\bibfnamefont{R.}~\bibnamefont{Wiesendanger}},
  \bibinfo{journal}{Science} \textbf{\bibinfo{volume}{341}},
  \bibinfo{pages}{636} (\bibinfo{year}{2013}), ISSN \bibinfo{issn}{0036-8075,
  1095-9203}, \bibinfo{note}{{PMID:} 23929977}.

\bibitem[{\citenamefont{Tchoe and Han}(2012)}]{tchoe_skyrmion_2012}
\bibinfo{author}{\bibfnamefont{Y.}~\bibnamefont{Tchoe}} \bibnamefont{and}
  \bibinfo{author}{\bibfnamefont{J.~H.} \bibnamefont{Han}},
  \bibinfo{journal}{Physical Review B} \textbf{\bibinfo{volume}{85}},
  \bibinfo{pages}{174416} (\bibinfo{year}{2012}).

\bibitem[{\citenamefont{Finazzi et~al.}(2013)\citenamefont{Finazzi, Savoini,
  Khorsand, Tsukamoto, Itoh, Du{\`o}, Kirilyuk, Rasing, and
  Ezawa}}]{finazzi_laser-induced_2013}
\bibinfo{author}{\bibfnamefont{M.}~\bibnamefont{Finazzi}},
  \bibinfo{author}{\bibfnamefont{M.}~\bibnamefont{Savoini}},
  \bibinfo{author}{\bibfnamefont{A.~R.} \bibnamefont{Khorsand}},
  \bibinfo{author}{\bibfnamefont{A.}~\bibnamefont{Tsukamoto}},
  \bibinfo{author}{\bibfnamefont{A.}~\bibnamefont{Itoh}},
  \bibinfo{author}{\bibfnamefont{L.}~\bibnamefont{Du{\`o}}},
  \bibinfo{author}{\bibfnamefont{A.}~\bibnamefont{Kirilyuk}},
  \bibinfo{author}{\bibfnamefont{T.}~\bibnamefont{Rasing}}, \bibnamefont{and}
  \bibinfo{author}{\bibfnamefont{M.}~\bibnamefont{Ezawa}},
  \bibinfo{journal}{Physical Review Letters} \textbf{\bibinfo{volume}{110}},
  \bibinfo{pages}{177205} (\bibinfo{year}{2013}).

\bibitem[{\citenamefont{Skyrme}(1961)}]{skyrme_non-linear_1961}
\bibinfo{author}{\bibfnamefont{T.~H.~R.} \bibnamefont{Skyrme}},
  \bibinfo{journal}{Proceedings of the Royal Society of London. Series A.
  Mathematical and Physical Sciences} \textbf{\bibinfo{volume}{260}},
  \bibinfo{pages}{127} (\bibinfo{year}{1961}), ISSN \bibinfo{issn}{1364-5021,
  1471-2946}.

\bibitem[{\citenamefont{Rajaraman}(1987)}]{rajaraman_solitons_1987}
\bibinfo{author}{\bibfnamefont{R.}~\bibnamefont{Rajaraman}},
  \emph{\bibinfo{title}{Solitons and Instantons}}
  (\bibinfo{publisher}{North-Holland}, \bibinfo{address}{Amsterdam},
  \bibinfo{year}{1987}).

\bibitem[{\citenamefont{Berg and L{\"u}scher}(1981)}]{berg_definition_1981}
\bibinfo{author}{\bibfnamefont{B.}~\bibnamefont{Berg}} \bibnamefont{and}
  \bibinfo{author}{\bibfnamefont{M.}~\bibnamefont{L{\"u}scher}},
  \bibinfo{journal}{Nuclear Physics B} \textbf{\bibinfo{volume}{190}},
  \bibinfo{pages}{412} (\bibinfo{year}{1981}), ISSN \bibinfo{issn}{0550-3213}.

\bibitem[{\citenamefont{Slonczewski}(1996)}]{slonczewski_current-driven_1996}
\bibinfo{author}{\bibfnamefont{J.}~\bibnamefont{Slonczewski}},
  \bibinfo{journal}{Journal of Magnetism and Magnetic Materials}
  \textbf{\bibinfo{volume}{159}}, \bibinfo{pages}{L1} (\bibinfo{year}{1996}),
  ISSN \bibinfo{issn}{0304-8853}.

\bibitem[{\citenamefont{Behin-Aein et~al.}(2010)\citenamefont{Behin-Aein,
  Datta, Salahuddin, and Datta}}]{behin-aein_proposal_2010}
\bibinfo{author}{\bibfnamefont{B.}~\bibnamefont{Behin-Aein}},
  \bibinfo{author}{\bibfnamefont{D.}~\bibnamefont{Datta}},
  \bibinfo{author}{\bibfnamefont{S.}~\bibnamefont{Salahuddin}},
  \bibnamefont{and} \bibinfo{author}{\bibfnamefont{S.}~\bibnamefont{Datta}},
  \bibinfo{journal}{Nat Nano} \textbf{\bibinfo{volume}{5}},
  \bibinfo{pages}{266} (\bibinfo{year}{2010}), ISSN \bibinfo{issn}{1748-3387}.

\bibitem[{\citenamefont{Brataas et~al.}(2012)\citenamefont{Brataas, Kent, and
  Ohno}}]{brataas_current-induced_2012}
\bibinfo{author}{\bibfnamefont{A.}~\bibnamefont{Brataas}},
  \bibinfo{author}{\bibfnamefont{A.~D.} \bibnamefont{Kent}}, \bibnamefont{and}
  \bibinfo{author}{\bibfnamefont{H.}~\bibnamefont{Ohno}}, \bibinfo{journal}{Nat
  Mater} \textbf{\bibinfo{volume}{11}}, \bibinfo{pages}{372}
  (\bibinfo{year}{2012}), ISSN \bibinfo{issn}{1476-1122}.

\bibitem[{\citenamefont{Karhu et~al.}(2012)\citenamefont{Karhu, R\"o\ss{}ler,
  Bogdanov, Kahwaji, Kirby, Fritzsche, Robertson, Majkrzak, and
  Monchesky}}]{karhu_chiral_2012}
\bibinfo{author}{\bibfnamefont{E.~A.} \bibnamefont{Karhu}},
  \bibinfo{author}{\bibfnamefont{U.~K.} \bibnamefont{R\"o\ss{}ler}},
  \bibinfo{author}{\bibfnamefont{A.~N.} \bibnamefont{Bogdanov}},
  \bibinfo{author}{\bibfnamefont{S.}~\bibnamefont{Kahwaji}},
  \bibinfo{author}{\bibfnamefont{B.~J.} \bibnamefont{Kirby}},
  \bibinfo{author}{\bibfnamefont{H.}~\bibnamefont{Fritzsche}},
  \bibinfo{author}{\bibfnamefont{M.~D.} \bibnamefont{Robertson}},
  \bibinfo{author}{\bibfnamefont{C.~F.} \bibnamefont{Majkrzak}},
  \bibnamefont{and} \bibinfo{author}{\bibfnamefont{T.~L.}
  \bibnamefont{Monchesky}}, \bibinfo{journal}{Phys. Rev. B}
  \textbf{\bibinfo{volume}{85}}, \bibinfo{pages}{094429}
  (\bibinfo{year}{2012}).

\bibitem[{\citenamefont{Huang and Chien}(2012)}]{huang_extended_2012}
\bibinfo{author}{\bibfnamefont{S.~X.} \bibnamefont{Huang}} \bibnamefont{and}
  \bibinfo{author}{\bibfnamefont{C.~L.} \bibnamefont{Chien}},
  \bibinfo{journal}{Phys. Rev. Lett.} \textbf{\bibinfo{volume}{108}},
  \bibinfo{pages}{267201} (\bibinfo{year}{2012}).

\bibitem[{\citenamefont{Garc{\'i}a-Palacios and
  L{\'a}zaro}(1998)}]{garcia-palacios_langevin-dynamics_1998}
\bibinfo{author}{\bibfnamefont{J.~L.} \bibnamefont{Garc{\'i}a-Palacios}}
  \bibnamefont{and} \bibinfo{author}{\bibfnamefont{F.~J.}
  \bibnamefont{L{\'a}zaro}}, \bibinfo{journal}{Physical Review B}
  \textbf{\bibinfo{volume}{58}}, \bibinfo{pages}{14937} (\bibinfo{year}{1998}).

\end{thebibliography}

\end{document}